# Isospin Separation of Hidden Non-Abelian Monopole

Le Van Hoang, Nguyen Thanh Son

Department of Physics, HCMC Education University,
280 An Duong Vuong, District 5, HCM City, Viet Nam,
Email: hoanglv@hcm.fpt.vn

**Abstract**: The scheme of isospin separation is suggested for the equation describing the five-dimensional 'charge-dyon' system in a non-Abelian SU(2) model. As a result, we obtain the Schrödinger equation for 'bare' particle, moving in Coulomb potential plus potential of five-dimensional Dirac monopole.

FACS: 0365 – w, Fd

In [1], we have generalized the Hurwitz transformation by introducing three Euler angles considered as new variables. With the use of this generalization, the authors of [2] interpret an eight-dimensional isotropic harmonic oscillator as hidden non-Abelian five-dimensional monopole (see, also, [3]). Indeed, after the Hurwitz transformation [4-7], the Schrödinger equation for a harmonic oscillator in eight-dimensional space leads to the following equation:

$$\left\{ \frac{1}{2}\left( -i\frac{\partial}{\partial x_\lambda} + B_{\lambda k}(\mathbf{r})\,\hat{Q}_k \right)^2 + \frac{1}{2r^2}\hat{Q}^2 - \frac{Z}{r} \right\} \psi(\mathbf{r},\phi) = E\psi(\mathbf{r},\phi), \qquad (1)$$

which describes a five-dimensional system "charge-dyon" in the non-Abelian SU(2) model. Here, in (1) the triplet of potential vectors $B_{\lambda k}(\mathbf{r})$:

$$B_{\lambda 1} = \frac{1}{r(r+x_5)}(x_2, -x_1, -x_4, x_3, 0),$$

$$B_{\lambda 2} = \frac{1}{r(r+x_5)}(-x_4, x_3, -x_2, x_1, 0), \qquad (2)$$

$$B_{\lambda 3} = \frac{1}{r(r+x_5)}(x_3, x_4, -x_1, -x_2, 0).$$

Meanwhile, three operators $\hat{Q}_1, \hat{Q}_2, \hat{Q}_3$ have the form:

$$\hat{Q}_1 = -i\frac{\partial}{\partial \phi_2},$$

$$\hat{Q}_2 = -i\frac{\sin\phi_2}{\sin\phi_3}\frac{\partial}{\partial \phi_1} + i\frac{\sin\phi_2}{\tan\phi_3}\frac{\partial}{\partial \phi_2} - i\cos\phi_2 \frac{\partial}{\partial \phi_3},$$

$$\hat{Q}_3 = -i\frac{\cos\phi_2}{\sin\phi_3}\frac{\partial}{\partial \phi_1} + i\frac{\cos\phi_2}{\tan\phi_3}\frac{\partial}{\partial \phi_2} + i\sin\phi_2 \frac{\partial}{\partial \phi_3},$$

and, consequently, satisfy the commutation relation:

$$\left[\hat{Q}_j, \hat{Q}_k\right] = i\,\varepsilon_{jkh}\,\hat{Q}_h. \qquad (3)$$

Now, we will show that for the equation (1) we can separate the dependence of angles $\phi_1, \phi_2, \phi_3$ to obtain the Schrödinger equation in five-dimensional space describing the hydrogen-like atom in 'electromagnetic' vector potential $A_\lambda(\mathbf{r})$. Moreover, the explicit form of $A_\lambda(\mathbf{r})$ obtained below allows us to interpret it as generalized Dirac monopole potential. This is the main result of present paper. In our notation the repeating of indices means summation over them and we use the unit system that $\hbar = m = c = e = 1$.

Because the variables $\phi_1, \phi_2, \phi_3$ according to (3) can be considered as three Euler angles with the generators of rotor $\hat{Q}_1, \hat{Q}_2, \hat{Q}_3$ correspondingly, we can construct also operators $\hat{T}_1, \hat{T}_2, \hat{T}_3$ with the following properties:

$$\left[\hat{T}_j, \hat{T}_k\right] = -i\,\varepsilon_{jkh}\,\hat{T}_h, \qquad \left[\hat{T}_j, \hat{Q}_k\right] = 0, \qquad \hat{T}_k\hat{T}_k = \hat{Q}_j\hat{Q}_j = \hat{Q}^2. \qquad (4)$$

For the physical meaning of the operators $\hat{T}_1, \hat{T}_2, \hat{T}_3$, one can see, for example, in the book [8]. Hereby we use the fact that operators $\hat{Q}_1, \hat{Q}_2, \hat{Q}_3$ coincide with $\hat{T}_1, \hat{T}_2, \hat{T}_3$ after the change:

$$\phi_1 \rightleftarrows \phi_2, \quad \phi_3 \to -\phi_3$$

to have:

$$\hat{T}_1 = -i\frac{\partial}{\partial \phi_1},$$

$$\hat{T}_2 = -i\frac{\cos\phi_1}{\tan\phi_3}\frac{\partial}{\partial \phi_1} + i\frac{\cos\phi_1}{\sin\phi_3}\frac{\partial}{\partial \phi_2} + i\sin\phi_1\frac{\partial}{\partial \phi_3},$$

$$\hat{T}_3 = -i\frac{\sin\phi_1}{\tan\phi_3}\frac{\partial}{\partial \phi_1} + i\frac{\sin\phi_1}{\sin\phi_3}\frac{\partial}{\partial \phi_2} + i\cos\phi_1\frac{\partial}{\partial \phi_3}.$$

Now let us separate the angles $\phi_1, \phi_2, \phi_3$ from equation (1). Noting that the Hamiltonian in (1) commutes with all of operators $\hat{Q}_1, \hat{Q}_2, \hat{Q}_3, \hat{T}_1, \hat{T}_2, \hat{T}_3$, we can build the angular part of wave-functions by requiring them to belong to eigen-functions of the operators $\hat{Q}^2, \hat{Q}_1, \hat{T}_1$:

$$\hat{Q}^2 \varphi^J_{q,p}(\phi) = J(J+1)\varphi^j_{q,p}(\phi),$$
$$\hat{Q}_1 \varphi^J_{q,p}(\phi) = q\,\varphi^J_{q,p}(\phi), \quad (5)$$
$$\hat{T}_1 \varphi^J_{q,p}(\phi) = p\,\varphi^J_{q,p}(\phi),$$

with $J = 0, 1, 2, \ldots$ ; $q, p = -J, -J+1, \ldots, J-1, J$ .

For the obvious form of the functions $\varphi^J_{q,p}(\phi)$, one can see in [8]. Here we just want to write some formulae to be used later on:

$$\hat{Q}_\pm \varphi^J_{q,p}(\phi) = \sqrt{(J \mp q)(J \pm q + 1)}\,\varphi^J_{q\pm 1,p}(\phi), \quad (6)$$

where $\hat{Q}_\pm = \hat{Q}_2 \pm i\hat{Q}_3$. For the angle separation process, we construct the wave-functions of equation (1) in the form:

$$\psi_{Jp}(\mathbf{r},\phi) = \Psi_{Jp}(\mathbf{r}) G_{Jp}(\mathbf{r},\phi), \quad (7)$$

with
$$G_{Jp}(\mathbf{r},\phi) = \sum_{q=-J}^{J} g_q(\mathbf{r}) \varphi^J_{pq}(\phi). \quad (8)$$

Using the notation:

$$\hat{a}_\lambda = B_{\lambda k}(\mathbf{r})\hat{Q}_k = B_{\lambda 1}(\mathbf{r})\hat{Q}_1 + B_{\lambda +}(\mathbf{r})\hat{Q}_+ + B_{\lambda -}(\mathbf{r})\hat{Q}_-$$

where $B_{\lambda +}(\mathbf{r}) = \frac{1}{2}(B_{\lambda 2}(\mathbf{r}) - iB_{\lambda 3}(\mathbf{r}))$, $B_{\lambda -}(\mathbf{r}) = \frac{1}{2}(B_{\lambda 2}(\mathbf{r}) + iB_{\lambda 3}(\mathbf{r}))$

we choose the coefficients $g_q(\mathbf{r})$ in the way that the angular part of wave-function (7) satisfies the equation:

$$\hat{a}_\lambda G_{Jp}(\mathbf{r},\phi) = A_\lambda(\mathbf{r}) G_{Jp}(\mathbf{r},\phi) . \tag{9}$$

Besides, from (5) we have two other equations for the angular part:

$$\hat{Q}^2 G_{Jp}(\mathbf{r},\phi) = J(J+1) G_{Jp}(\mathbf{r},\phi), \tag{10}$$

$$\hat{T}_1 G_{Jp}(\mathbf{r},\phi) = p G_{Jp}(\mathbf{r},\phi) . \tag{11}$$

Now let us resolve equation (9) by substituting (8) into it and then using formula (6). As a result, for each term $\lambda = 1, 2, ..., 5$ we obtain a system of $2J+1$ linear uniform equations:

$$\sqrt{(J-q+1)(J+q)}B_{\lambda +}(\mathbf{r})g_{q-1}(\mathbf{r}) - \left[A_\lambda(\mathbf{r}) - qB_{\lambda 1}(\mathbf{r})\right]g_q(\mathbf{r})$$

$$+ \sqrt{(J+q+1)(J-q)}B_{\lambda -}(\mathbf{r})g_{q+1}(\mathbf{r}) = 0 \tag{12}$$

for $q = -J, -J+1, -J+2, ..., J-1, J$ . Since system of equations (12) is uniform, it has nontrivial solution only in the case when the determinant of the following matrix:

$$h_\lambda = \begin{pmatrix} -A_\lambda - JB_{\lambda 1} & \sqrt{2J}B_{\lambda -} & 0 & \cdots & 0 \\ \sqrt{2J}B_{\lambda +} & -A_\lambda - (J-1)B_{\lambda 1} & \sqrt{2(2J-1)}B_{\lambda -} & \cdots & 0 \\ 0 & \sqrt{2(2J-1)}B_{\lambda +} & -A_\lambda - (J-2)B_{\lambda 1} & \cdots & 0 \\ \cdots & \cdots & \cdots & \cdots & \sqrt{2J}B_{\lambda -} \\ 0 & 0 & 0 & \sqrt{2J}B_{\lambda +} & -A_\lambda + JB_{\lambda 1} \end{pmatrix}$$

is vanished. Noticing that only the following elements of matrix $h_\lambda$ are different from zero

$$(h_\lambda)_{k,k-1} = \sqrt{(2J+2-k)(k-1)} B_{\lambda+}(\mathbf{r}),$$

$$(h_\lambda)_{kk} = -A_\lambda(\mathbf{r}) - (J+1-k)B_{\lambda 1}(\mathbf{r}),$$

$$(h_\lambda)_{k,k+1} = \sqrt{k(2J+1-k)} B_{\lambda-}(\mathbf{r}) \ , k = 1, 2, \ldots, 2J+1$$

we can use Mathematica [9] to calculate determinant of matrix $h_\lambda$ and obtain:

$$\det(h_\lambda) = A_\lambda(\mathbf{r}) \left[ A_\lambda^2(\mathbf{r}) - \left(B_{\lambda 1}^2 + B_{\lambda 2}^2 + B_{\lambda 3}^2\right)\right]\left[A_\lambda^2(\mathbf{r}) - 4\left(B_{\lambda 1}^2 + B_{\lambda 2}^2 + B_{\lambda 3}^2\right)\right]\cdots$$
$$\cdots\left[A_\lambda^2(\mathbf{r}) - J^2\left(B_{\lambda 1}^2 + B_{\lambda 2}^2 + B_{\lambda 3}^2\right)\right].$$

The condition $\det(h_\lambda) = 0$ leads to an algebraic equation with order of $2J+1$ for the root $A_\lambda(\mathbf{r})$ which is easy to obtain in the form:

$$A_\lambda(\mathbf{r}) = j_\lambda \sqrt{B_{\lambda 1}^2 + B_{\lambda 2}^2 + B_{\lambda 3}^2}$$

with $j_\lambda = -J, -J+1, \ldots, J-1, J$. Taking into account (2) we have:

$$A_1(\mathbf{r}) = j_1 \frac{\sqrt{x_2^2 + x_3^2 + x_4^2}}{r(r+x_5)}, \quad A_2(\mathbf{r}) = j_2 \frac{\sqrt{x_1^2 + x_3^2 + x_4^2}}{r(r+x_5)},$$

$$A_3(\mathbf{r}) = j_3 \frac{\sqrt{x_1^2 + x_2^2 + x_4^2}}{r(r+x_5)}, \quad A_4(\mathbf{r}) = j_4 \frac{\sqrt{x_1^2 + x_2^2 + x_3^2}}{r(r+x_5)}, \quad A_5(\mathbf{r}) = 0.$$

(13)

Putting the found $A_\lambda(\mathbf{r})$ into equations (12) we can find $2J+1$ coefficients $g_q(\mathbf{r})$ for each term $\lambda$. It means that the angular part of wave-functions $G_{Jp}(\mathbf{r}, \phi)$ can obviously be constructed. Now let us put the wave functions (8) into equation (1), taking into account the equations (9), (10), (11), we obtain an equation for the motion in 'physical' $\mathbf{r}$ – space:

6                          Le Van Hoang, Nguyen Thanh Son

$$\left\{\frac{1}{2}\left(-i\frac{\partial}{\partial x_\lambda}+A_\lambda(\mathbf{r})\right)^2+\frac{J(J+1)}{2r^2}-\frac{Z}{r}\right\}\Psi_{Jp}(\mathbf{r})=E\,\Psi_{Jp}(\mathbf{r})\ . \qquad (14)$$

Consequently, we have successfully separated the 'extra' variables from the equation (1), describing the non-Abelian monopole. Because equation (1) and the Schrödinger equation for eight-dimensional harmonic oscillator are dual to each other [2], we can say that the separation process built in the present paper allows us to connect the harmonic oscillator problem with the problem of five-dimensional hydrogen-like atom in vector potential of Dirac monopole. The equation (1) principally describes the isospin particle moving in the Coulomb potential and interacting with the vector potential (2). Here, three generators of extra variables characterize an internal property of particle (isospin). Thus, the construction of equation (14) means that we have successfully extracted the isospin dependence from the equation and, as a result, received the equation described motion of a 'bare' particle without isospin.